\documentclass[conference,twocolumn,letterpaper,10pt]{IEEEtran}

\usepackage{amsmath,amssymb,amsfonts}
\usepackage[ruled,vlined]{algorithm2e}% http://ctan.org/pkg/algorithms
\usepackage{algorithmic}
\usepackage{bm}
\usepackage{braket}
\usepackage{cite}
\usepackage{color}
\usepackage{diagbox}
\usepackage{epstopdf}
\usepackage{graphicx}
\usepackage{multirow}
\usepackage{overpic}
\usepackage{pdfsync}
\usepackage{setspace}
\usepackage{soul}
\usepackage{subcaption}
\usepackage{tabularx}
\usepackage{verbatim}
\usepackage[usenames,dvipsnames,svgnames,table]{xcolor}

\renewcommand{\baselinestretch}{0.99}

\begin{document}

\newtheorem{condition}{Condition}
\allowdisplaybreaks 
%\singlespacing
%\onehalfspacintg
%\doublespacing
%\renewcommand{\sfdefault}{phv}
%\theoremstyle{definition}
%\theoremstyle{remark}
%\theoremstyle{plain}
\newtheorem{assu}{Assumption}
\newtheorem{cor}{Corollary}
\newtheorem{defn}{Definition}
\newtheorem{lem}{Lemma}
\newtheorem{prop}{Proposition}
\newtheorem{thm}{Theorem}
\newcommand{\R}{\mathbb{R}}
\newcommand{\Z}{\mathbb{Z}}
\newcommand{\N}{\mathbb{N}}
\newcommand{\1}[1]{1_{[#1]}}
\newcommand{\dd}{\mathrm{d}}
\newcommand{\MMT}{\mathsf{MMT}}
\newcommand{\bkref}[1] {(\ref{#1})}

\title{
Adaptive Robust Traffic Engineering\\in Software Defined Networks
}

%\author{
%\IEEEauthorblockN{double blind}
%}

\author{\IEEEauthorblockN{Davide Sanvito, Ilario Filippini, Antonio Capone	}

 \IEEEauthorblockA{Politecnico di Milano\\{\it name.surname}@polimi.it}
 \and
 \IEEEauthorblockN{Stefano Paris, Jeremie Leguay}
 \IEEEauthorblockA{France Research Center, Huawei Technologies Co. Ltd\\{\it name.surname}@huawei.com
% Boulogne Billancourt, France\\
% Email: name.surname@huawei.com
}
 }
 
%\date{}
\maketitle

\newcommand{\comm}[1]{\textcolor{red}{\hl{\em$[$#1$]$}}}
\newcommand{\change}[1]{\textcolor{blue}{#1}}

%\newtheorem{theorem}{Theorem}[section]
%\newtheorem{lemma}{Lemma}[section]
%\newtheorem{proposition}{Proposition}[section]
%\newtheorem{corollary}{Corollary}[section]

%-------------------------------------------
% Abstract (with keywords)
%-------------------------------------------
\begin{abstract}

%Shortened version to meet the 200 words constraint
One of the key advantages of Software-Defined Networks (SDN) is the opportunity to integrate traffic engineering modules able to optimize network configuration according to traffic. Ideally, network should be dynamically reconfigured as traffic evolves, so as to achieve remarkable gains in the efficient use of resources with respect to traditional static approaches. Unfortunately, reconfigurations cannot be too frequent due to a number of reasons related to route stability, forwarding rules instantiation, individual flows dynamics, traffic monitoring overhead, etc.

In this paper, we focus on the fundamental problem of deciding whether, when and how to reconfigure the network during traffic evolution. We propose a new approach to cluster relevant points in the multi-dimensional traffic space taking into account similarities in optimal routing and not only in traffic values. Moreover, to provide more flexibility to the online decisions on when applying a reconfiguration, we allow some overlap between clusters that can guarantee a good-quality routing regardless of the transition instant.

We compare our algorithm with state-of-the-art approaches in realistic network scenarios. Results show that our method significantly reduces the number of reconfigurations with a negligible deviation of the network performance with respect to the continuous update of the network configuration.

\end{abstract}

%\begin{keywords}
%\textit{Software Defined Networking, Routing, Robust Optimization.}
%\end{keywords}

\pagestyle{plain}

%-------------------------------------------
% Introduction
%-------------------------------------------
\section{Introduction}
\label{introduction}

%%%%%%%%%%%%%%%%%%%%%%%%%%%%%%%%%%%%%%%%%%%%%%%
% 1. Context
%%%%%%%%%%%%%%%%%%%%%%%%%%%%%%%%%%%%%%%%%%%%%%%
% Traffic engineering (TE) is key and challenging
Traffic Engineering (TE)~\cite{Wang08} plays a crucial role for service providers since it permits to optimize network performance, reduce operational costs, and load balance the utilization of network resources. 
%Such an optimization of routing is challenging because of the dynamic nature of traffic induced by repeating patterns (e.g., diurnal demand fluctuation) and unpredictable events (e.g., flash crowds, denial of service attacks). In the practice, service providers always prefer to avoid frequent routing changes due to operational complexity and costs. They resort most of the times to network planning tools to re-optimize routing at a low frequency (e.g., every month) as they dread network outages caused by route update failures or cascading traffic changes difficult to predict. Unfortunately, the trade-off in avoiding routing changes is the significant overprovisioning capacity needed to accommodate any traffic variations while keeping the routing fixed.
However, the dynamic nature of the traffic due to ordinary daily fluctuations and unpredictable events stirs up the trade-off between optimality of the routing configuration and network reconfiguration rate.
%, which might result in route flapping. 
The traditional approach of service providers is to optimize the routing considering the "worst case" traffic condition so as to rarely reconfigure the network. The resulting overprovisioning leads to the underutilization of network resources.

% SDN will help to implement online TE solutions
% Nowadays, Software-Defined Network (SDN) architectures~\cite{kreutz2015software} give service providers the opportunity to update more frequently traffic engineering policies.
% SDN has unleashed the potential to compute routing at a powerful central controller
% and then reconfigure the network accordingly in real-time. It uses modern distributed computing technologies (e.g., consensus algorithms, transactional updates, failover) to gather and keep a global view of the network status in real-time and to reliably push consistent configuration updates to the network equipment. Sophisticated algorithms have been designed to efficiently implement TE in SDN~\cite{hong2013achieving, jain2013b4, istamp14}. As the network evolves new configurations are applied to devices to optimize network performance. 
% However, sudden and unpredictable system changes, like traffic variations and network failures, still pose a major challenge for these methods.
Software-Defined Networks (SDNs)~\cite{kreutz2015software} provide the needed flexibility to update more frequently TE policies. Having a global view of the network status, SDN controllers can integrate TE algorithms~\cite{hong2013achieving, jain2013b4, istamp14} to continuously optimize the network with an online twist. As the system evolves, new configurations are applied to the network equipment to optimize network performance.
However, sudden and unpredictable system changes, like traffic variations, network failures, and the uncontrolled rate of change, still pose a major challenge for these methods.

% Typical classes of TE solutions
The solutions that have been devised to cope with traffic variations can be broadly classified into three main classes: \textit{dynamic TE}, \textit{static TE}, and \textit{semi-static TE}. While different in the way they calculate network configurations, all techniques require the use of Traffic Matrices (TMs) periodically collected by a network monitoring tool.
\textit{Dynamic TE}, like \cite{jain2013b4, hong2013achieving, benson2011microte,roughan2003traffic}, uses such information to predict the next system state and compute the corresponding optimal routing configuration using linear programming~\cite{Murakami98} or fast approximation algorithms~\cite{Albrecht01}.
The accuracy of the prediction highly affects the optimality of the computed solution while frequent network reconfigurations result in control plane congestion due to the low speed of flow programming~\cite{jain2013b4} in hardware. 
%Methods have been proposed to reduce this burden by prioritizing~\cite{jin2014dynamic} or pre-filtering~\cite{paris2016controlling} network updates. But the computation of each routing solution does not consider any robustness considerations.
In contrast, \textit{static TE}, such as \emph{oblivious routing}~\cite{azar2003optimal} and robust routing~\cite{tabatabaee2007robust, kodialam2004efficient, wang2006cope}, monitors TMs over a long period of time and computes the TE configuration that minimizes the worst deviation with respect to the sequence of all optimal configurations. This class of TE policies keeps the network configuration stable, but it inevitably suffers from low optimality during most of the operational time.

% Shortcomings of current semi-static TE strategies
\textit{Semi-static TE} approaches such as~\cite{Casas08, zhang2005finding} combine both static and dynamic TE to approximate the optimal sequence of configurations with a limited set of routing solutions computed over clusters of TMs.
Clusters of TMs are formed either by statically dividing time in different intervals or by finding similarities in the traffic domain. 
However, the arbitrary splitting of the time domain results in significant performance loss when sharp traffic variations are temporally close. Similarly, using the same routing configuration for TMs that are close in the traffic domain (i.e., their entries have the same magnitude) but far in the time domain can lead to frequent network reconfigurations. 
%Although cluster-based approaches have the potential to optimize network performance using a limited set of routing configurations, several dimensions should be considered when building clusters. And also, these clusters should be built in such as way that the controller can decide when to reconfigure. 
%Oscillations between routing configurations  should be avoids and a reconfiguration should be carried out only when the system is evolving towards a new equilibrium/state. The solution proposed in this paper aims at solving these issues.
More importantly, the controller needs to decide whether and when to reconfigure. Transitory traffic fluctuations should be ignored to avoid system oscillations and the network should be reconfigured only when it is evolving towards a new state.

%%%%%%%%%%%%%%%%%%%%%%%%%%%%%%%%%%%%%%%%%%%%%%
% 2. Our proposal
%%%%%%%%%%%%%%%%%%%%%%%%%%%%%%%%%%%%%%%%%%%%%%
In this paper, we study the fundamental problem faced by SDN controllers of deciding whether, when and how to reconfigure the network after a traffic evolution.
To provide an answer we study and address the problem of building a set of robust routing configurations associated to clusters of TMs that overlaps in time, traffic and routing domains. 
Time overlap refers to the amount of time we are able to use a routing configuration even for TMs that are outside the associated cluster with minimal efficiency degradation.
Traffic overlap denotes the similarity in the traffic space of TMs within the same cluster, whereas routing overlap indicates how similar routing configurations associated to two different clusters are. 
%The approach somehow shares some similarities with cell partitioning in mobile systems where overlap facilitates mobility management, but with additional elements due to multi-dimensional traffic space and network topology that determines routing.{\color{red}\textbf{JL: I would remove the last sentence.}}
%\comm{IF(I was a little bit confused by the overlap in domains different from time. What about this proposal?) clusters of TMs that extend over time, traffic and routing domains. 
%Time domain refers to the minimum amount of time we should be able to use a routing configuration, even for TMs that are outside the associated cluster.
%Traffic domain denotes the similarity in the traffic space of TMs within the same cluster, whereas routing domain indicates how similar is the routing performance of TMs associated to the same cluster.}
% In order to achive a good trade-off between network stability and optimality, we present in this paper a method to build a set of robust routing configurations associated to clusters of traffic matrices that overlaps in time, traffic and routing domains.

Given the interplay between TM clusters and routing solutions, we decouple the problem into two subproblems, namely TM clustering and robust routing. 
To this aim, we propose Clustered Robust Routing (CRR), an iterative algorithm that achieves three objectives: 1) covering the entire TM space so that a feasible routing configuration is available for any traffic condition, 2) reducing the number of routing changes by creating a small set of robust routing configurations that can be used for a minimum duration each time one of them is applied, and 3) maintaining a minimum time overlap between adjacent clusters that can be exploited to decide whether to reconfigure the network.
We analyze our algorithm on a realistic network scenario and compare its performance against state of the art approaches of the three TE classes discussed above.

%%%%%%%%%%%%%%%%%%%%%%%%%%%%%%%%%%%%%%%%%%%%%%
% 3. Structure
%%%%%%%%%%%%%%%%%%%%%%%%%%%%%%%%%%%%%%%%%%%%%%
This paper is structured as follows:
Section~\ref{related_work} presents the related work.
Section~\ref{system_model} describes the system model and the assumptions we made in the formulation of our problem.
Section~\ref{crr_algorithm} presents the algorithm to build clustered robust configurations considering the time continuum, the traffic space and routing similarities. Numerical results are discussed in Section~\ref{results}.
Finally, concluding remarks are presented in Section~\ref{conclusion}.

%-------------------------------------------
% Related Work
%-------------------------------------------
\section{Related Work}
\label{related_work}

The simplicity of controlling SDNs has brought back to light problems like mitigating and scheduling network reconfigurations~\cite{holterbach2017swift,brandt2016consistent}, since service providers are concerned about possible network outages caused by failures of route updates or sudden traffic changes.
%However, the dynamic nature of the traffic induced by repeating patterns (e.g., diurnal demand fluctuation) and unpredictable events (e.g., flash crowds, denial of service attacks).
%In the practice, service providers always prefer to avoid frequent routing changes due to operational complexity and costs. They resort most of the times to network planning tools to re-optimize routing at a low frequency (e.g., every month) as they dread network outages caused by route update failures or cascading traffic changes difficult to predict. Unfortunately, the trade-off in avoiding routing changes is the significant overprovisioning capacity needed to accommodate any traffic variations while keeping the routing fixed.
The networking research community has developed three classes of techniques to handle traffic change: (i) \textit{dynamic TE}, which reconfigures the network each time a new event occurs, (ii) \textit{static TE}, which uses a single precomputed configuration that minimizes the worst deviation to the optimum, and (iii) \textit{semi-static TE}, which reconfigures the network at predefined time instants (e.g., twice per day at noon and midnight) to further improve network performance.
Examples of \textit{dynamic TE} includes methods like~\cite{jain2013b4,hong2013achieving,benson2011microte,roughan2003traffic}, where sophisticated techniques are used to compute the best network configuration any time the traffic changes. However, reconfiguring the network too frequently can affect its stability, since programming hardware equipment with new flow rules can take longer than the reconfiguration period~\cite{jain2013b4}, thus causing the overflow of flow rules.
Methods that reduce this burden have been proposed by prioritizing~\cite{jin2014dynamic} or pre-filtering~\cite{paris2016controlling} network updates. Nonetheless, the computation of each routing solution is not robust against prediction errors on the next TM.

% \begin{figure}[t]
% 	\centering
% 	\includegraphics[width=0.8\linewidth]{crr_dim.png}
% 	\caption{\small{Dimensions considered for the clustering of Traffic Matrices and computation of routing configurations. The traffic dimension is in reality multidimensional (one dimension for each OD flow).}}
% 	\label{fig:crr_dim}
% \end{figure}

One of the first techniques of \textit{static TE} is oblivious routing~\cite{azar2003optimal, applegate2003making}, and its recent extension called \emph{valiant routing}~\cite{Zhang-Shen2010}, which randomly selects paths to connect source-destination pairs using a small subset of preselected intermediate nodes.
%without using any knowledge of the traffic. 
Being totally oblivious to any traffic information, oblivious routing shows high performance loss as the network size grows.
Exploring a partial knowledge of the traffic can reduce the performance loss. For example, COPE~\cite{wang2006cope} considers only the most likely TMs for computing the optimal configuration and add a penalty term to avoid large deviation for less probable TMs. The technique proposed in~\cite{casas2009robust} expands the most likely polytope by including TMs of normal operations in the direction of a predicted anomaly.
The method proposed in~\cite{tabatabaee2007robust} introduces different models for traffic uncertainness by expressing the maximum load that can be expected over a link in the pipe model or an upper bound on the traffic originating from a source node and directed to a destination node in the hose model.

\textit{Semi-static TE}~\cite{Casas08, ben2011robust, zhang2005finding} provides a limited set of routing configurations with guaranteed performance loss. These works divide the TM polytope in two subsets according to the time dimension and compute a robust routing for each subset. While representing a first attempt to split the TM domain in multiple parts, these works present several limitations: (i) the slicing direction is arbitrary, (ii) the number of created subsets is limited, and (iii) the partition is performed either in the traffic domain or in the time domain.

% *** NEW ***
Although semi-static TE approaches have the potential to optimize network performance using a limited set of routing configurations, traffic, time, and routing spaces/dimensions should be jointly considered when building clusters in order to avoid oscillations between routing configurations when TMs are close in the traffic space but far in the time dimension.
Furthermore, clusters should not be sharply separated, since instantaneous routing changes are impossible even in SDNs.
% And also, these clusters should be built in such as way that the controller can decide when to reconfigure. 
% Oscillations between routing configurations  should be avoids and a reconfiguration should be carried out only when the system is evolving towards a new equilibrium/state. The solution proposed in this paper aims at solving these issues.
This work is a first attempt to address these problems and decide the best trade-off between reconfiguration rate and optimality of  routing.
% *** OLD ***
% {\em Main contributions} 
% The main contributions of this paper are listed hereafter:
% \begin{enumerate}
% \item Formulation of a robust routing problem to fully exploit the time continuum, as well as the traffic and routing similarities of TM observed in the past.
% \item Proposition of Clustered Robust Routing (CRR), an iterative algorithm to create a small set of robust routing configurations covering the entire traffic matrix space and which can be applied for a minimum amount of time.
% \item Evaluation of our algorithm on a realistic network scenario in comparison to state of the art approaches.
% \end{enumerate}

%-------------------------------------------
% Problem description, assumptions
%-------------------------------------------
\section{System Model}
\label{system_model}

In this section, we present the traffic and routing system models that we consider in the design of our CRR algorithm.

We consider a system composed of two main stages: (i) an offline stage where we group TMs into clusters and compute robust routing configurations over these clusters, and (ii) an online stage where we track the traffic evolution and reconfigure the network accordingly.
The target is to minimize the Maximum Link Utilization (MLU) over time, which is motivated in the domain of datacenter interconnection and enterprise networks, where the goal is to minimize the network congestion. 
%since several studies have demonstrated the strong relationship between MLU and network congestion. 

We model the network infrastructure as an undirected graph $G=(\mathcal{N}, \mathcal{L})$, where $\mathcal{N}$ represents the set of network nodes and $\mathcal{L}$ models the set of links $e = (i,j)$, connecting network nodes $i,j \in \mathcal{N}$.
Each link $e \in \mathcal{L}$ has a limited capacity $c_{ij}$ that represents the maximum amount of traffic that the link can transmit.
%A demand $k \in \mathcal{K}$, also known as Origin-Destination (OD) flow, is identified by a source-destination pair $(s_{k},t_{k}) \in \mathcal{N}^2$, and a variable amount of traffic $d_{k}(t)$ that has to be transmitted from $s_{k}$~to~$t_{k}$.
The set of active demands, also known as Origin-Destination (OD) flows, that need to be routed through the network, is represented as a Traffic Matrix (TM): a $\vert\mathcal{N}\vert\times\vert\mathcal{N}\vert$ matrix $T=\left[t_{ij}\right]$ where each element $t_{ij}$ denotes the amount of traffic transmitted from source node~$i$~to destination node~$j$.
Since the traffic evolves during time, we consider a dynamic TM $T(\tau)=\left[t_{ij}^{\tau}\right]$, where $\tau$ denotes the time dimension. We assume that time is discretized and we have $M$ samples of the TM (i.e., $\tau=1,...,M$).
To simplify the notation, TM are usually represented as a $\vert\mathcal{N}\vert^{2}\times~1$ demand vector $D(\tau)=\left[d_{h}^{\tau}\right]$ where each element $d_{h}^{\tau}$ unequivocally corresponds to an element $t_{ij}^{\tau}$ of the TM.

\begin{figure}[t]
	\centering
\includegraphics[width=0.9\linewidth]{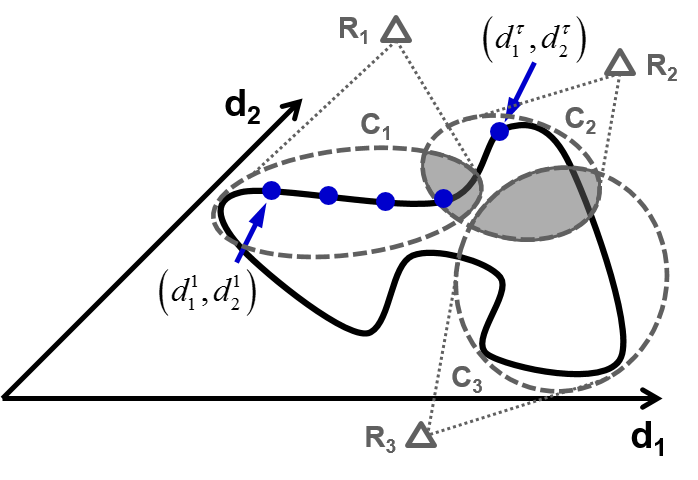}
	\caption{\small{Clustering of TMs. The solid black line represents the evolution of the two OD flows in the TM. Each blue point represents a sampled TM. Dashed ellipsoids denote the clusters of TMs, whereas triangles identify the corresponding routing configurations.}}
    \vspace{-1.5em}
	\label{fig:crr_idea}
\end{figure}

\textbf{Offline cluster maintenance.} Fig.~\ref{fig:crr_idea} graphically illustrates the time evolution of a TM composed by only two demands, $d_{1}$ and $d_{2}$. 
The solid line represents the continuous evolution of the TM, whereas solid dots corresponds to periodic samples measured by a traffic monitoring system. As illustrated in the figure, an offline stage splits the TM domain into $N$ clusters, denoted as $C_{i}$, and computes for each subset of TMs a routing configuration $R_{i}$, which is robust against any possible traffic variation within the cluster $C_{i}$. To avoid oscillations between routing configurations, a cluster $C_i$ is built with a minimum time length $L$ that results in a minimum utilization of the same routing configuration $C_i$.
Furthermore, a temporal overlap $O$ (the gray intersection in Fig.~\ref{fig:crr_idea}) is imposed between two adjacent clusters $C_{i}$ and $C_{j}$ to guarantee the feasibility of the corresponding robust routing configurations $R_{i}$ and $R_{j}$ outside their clusters. The overlap $O$ leaves enough time to the SDN controller to decide whether to reconfigure the network and prefetch the new routing configuration in the switches.

\textbf{Online cluster activation.} 
The different precomputed routing configurations are then activated in an online manner by the SDN controller which follows the evolution of the traffic matrix. By receiving an estimate of actual traffic conditions (and possibly a short term prediction) from the monitoring system, it decides whether to fetch and activate a better robust routing configuration in switches. 

Clearly, the performance of this approach depends on the size and the number of clusters. Many small clusters result in a high reconfiguration frequency, which may harms the network behavior itself.
In contrast, too few clusters will provide a low gain over static-TE solutions (e.g., oblivious routing). 
This approach will always lead to a better performance than the single-design case, because the network is no longer forced to always support the worst-case traffic demand. Indeed, the correct solution will be applied when the corresponding worst-case (among the possibly many that lie in different regions) appears. 

%The definition of the boundary between regions is itself a decision that can highly affect the final network performance. 
Furthermore, a sharp boundary between clusters that are adjacent in the time domain requires an instantaneous reconfiguration of the network.
For a smooth network reconfiguration when the TM enters into a new cluster, we compute clusters with a minimum time overlap, represented by the intersection of two clusters in Fig.~\ref{fig:crr_idea}. This leaves enough time to the SDN controller to decide whether to reconfigure the network and pre-fetch the new routing configuration into switches.

In the next section, we show how our algorithm, Clustered Robut Routing (CRR), solves the problem of achieving a good trade-off between routing stability and optimality by maintaining a set of  routing configurations for overlapping clusters of similar traffic matrices.

 Table~\ref{tab:notation_input} summarizes the notation used throughout the paper.
% %
\begin{table}[t!]
{\small
\centering
\begin{tabular}{|c|l|}
	\hline
	\textbf{Parameter} & \textbf{Description} \\
	\hline
	$\mathcal{N}$ & Nodes (network devices). \\
	\hline
	$\mathcal{L}$ & Edges (network links). \\
	\hline
	$c_{ij}$ & edge capacity (in capacity units) $(i,j) \in \mathcal{L}$. \\
	\hline
	$d_{h}^{\tau}$ & rate of OD demand $d$ measured at time $\tau$. \\
	\hline
    $N$ & number of robust routing configurations. \\
    \hline
    $M$ & number of traffic matrices. \\
    \hline
    $L$ & minimum holding time of a routing configuration. \\
    \hline
    $O$ & temporal overlap between two adjacent clusters.\\
    \hline
\end{tabular}
}
\caption{Input parameters of our CRR algorithm.}
\vspace{-1.5em}
\label{tab:notation_input}
\end{table}
\normalsize

%-------------------------------------------
% Clustered Robust Routing Algorithm
%-------------------------------------------
\section{Clustered Robust Routing}
\label{crr_algorithm}

The CRR algorithm is implemented as a module of the network controller. It takes as input a set of TMs representative of the period in which robust routing configurations should be designed. These TMs can be obtained in several ways: they can be measurements from past network conditions, or the outcome of a TM prediction module, or even synthetically generated. For the sake of clarity, we neglect the effect of prediction errors in the description of the algorithm, however, we investigate the impact of inaccurate TMs within numerical results. The impact is indeed rather limited for realistic error values because the clustering generates intrinsically robust solutions. Each TM describes the expected traffic conditions at specific time instants. Therefore, TMs can be temporally ordered and the set of TM IDs can be used as time axis. The result of the algorithm is a set of routing configurations (RCs, denoted as $R_{i}$ in Fig.~\ref{fig:crr_idea}) and the corresponding clusters of TMs ($C_i$ in Fig.~\ref{fig:crr_idea}). Each RC will be activated in the network as soon as the traffic enters the corresponding cluster. 

\subsection{Requirements for CRR} 

The CRR algorithm, shown in Fig.~\ref{fig:crr_algo}, consists in an iterative clustering and routing process relying on four main points:

\paragraph{Routing-based clustering} A TM clustering approach based on the similarity among the OD demands of each TM can be strongly inefficient. Indeed, since network's links have limited capacity, good quality routes can substantially differ for TMs with similar demands. Since the way traffic is balanced over the network is not captured by the unique RC used to route the TM cluster, which is based on demand values, it may lead to high congestion for some TMs.

Things do not improve even if TMs with similar optimum routing are grouped together to generate a good unique cluster RC. Indeed, as we will show in Sec.\ref{results}, this does not provide the best results. Due to scenario symmetries, different RCs can provide the same congestion, therefore clustering on the mere basis of RC topology can be largely suboptimal. Indeed, since the number of desired clusters in a solution is usually limited, this approach may waste clusters to separate TMs with different optimum routings, which could be equivalently well routed by another unique RC.

In order to better include the routing effects in the cluster selection, we need to consider the ultimate effect of the routing, that is the network congestion resulting from applying a given RC to a given TM. Only TMs that are characterized by a small congestion with the same RC must be grouped together into the cluster associated to the specific RC.

\paragraph{In-cluster robust routing} Although clustering is based on routing (i.e., it groups TMs having a similar congestion with a specific RC), the RC design cannot be strongly customized on a specific TM. Indeed, in practice we have to deal with deviation from the TM input set or even traffic anomalies. Therefore, we need a robust routing solution to cope with the demand variations of the clustered TMs. 

\begin{figure}[t]
	\centering
	\includegraphics[width=0.6\linewidth]{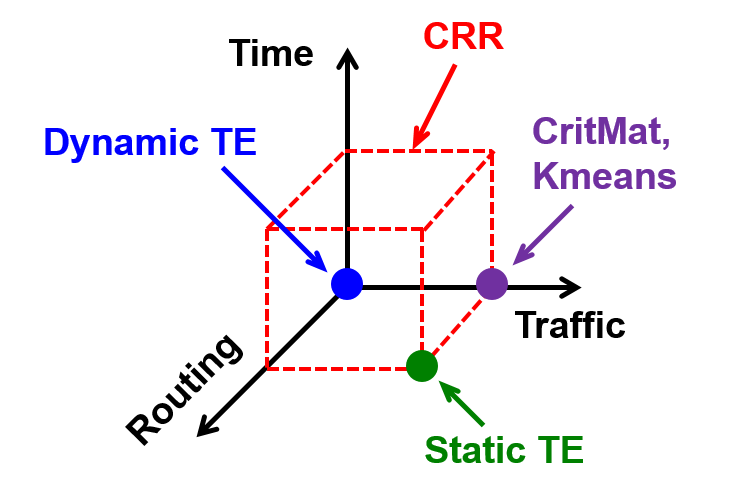}
	\caption{\small{Dimensions considered for the clustering of Traffic Matrices and computation of routing configurations. The traffic dimension is in reality multidimensional (one dimension for each OD flow).}}
    \vspace{-1em}
	\label{fig:crr_dim}
\end{figure}

There are several approaches to robust routing in literature. The most straightforward solution is to consider the convex hull of all the discrete TMs and design the routing for the worst case in this continuous set. However, this approach has a number of drawbacks as the outcome may be strongly affected by a particular combination of demands that can be very rare in practice, thus producing an excessively conservative RC. In addition, the optimization process can be quite complicated \cite{applegate2003making,ben2005routing}. Since we assume a set of representative TM to be available, which represent the most likely or most important network conditions for the routing optimization process, we prefer to rely on a discrete space and adopt a multi-TM robust optimization approach, like those in \cite{zhang2005optimal}. A possible alternative when significant anomalies come into play is the approach presented in \cite{wang2006cope}, in which the optimization process still focuses on the set of most representative TMs, while a bounded penalty gap is guaranteed over the remaining traffic domain, thus also in case of anomalies. However, a complete analysis of the anomaly management is out of the scope of this paper.

\paragraph{Routing configuration holding time} Although SDN provides flexible and efficient tools to dynamically change network routes, we must pay attention not to change the network configuration too rapidly, incurring into route flapping problems. Therefore, the CRR algorithm includes for each activated RC a minimum holding time before reconfiguring. If the set of considered TMs is a uniform sampling of the expected traffic conditions, the minimum holding time is equivalently described by a minimum number of consecutive TMs in each cluster.

This feature brings in a new dimension in the clustering problem by adding the time dimension together with routing and traffic. Figure~\ref{fig:crr_dim} illustrates graphically the design space of our algorithm and how the other approaches locate with respect to our proposal. Exploiting the time continuum, as well as the traffic and routing, it allows to improve the network performance and at the same time to reduce the number of network reconfigurations.

\paragraph{Adjacent clusters overlap}
The transition of traffic conditions from those described by the current cluster to those of a new cluster must be carefully addressed to maintain a good routing quality. Although the technical route update process has been thoroughly studied and several SDN-based consistent update schemes are available~\cite{reitblatt2011consistent, wang2016cupid}, a further fundamental issue is to decide when this update should occur. Different algorithms can be implemented to decide the best switching point depending on the past, current and predicted traffic behavior, considering anticipatory networking aspects as well. However, the common aspect among them will be an unavoidable uncertainness about the time to switch. Therefore, considering an overlap among adjacent clusters is important, because it guarantees a graceful transition between them. This means that RCs of adjacent clusters will be reasonably good with the TMs that are expected to be close to a route transition. This helps algorithms not to be too much penalized from suboptimal decisions.

Besides making reconfigurations robust to prediction errors, the time overlap facilitates the network reaction to traffic changes. SDN switches can store several RCs, the active one and a set of RCs potentially useful in the immediate future, according to traffic predictions. The time overlap allows to pre-fetch next RCs before reaching the cluster boundary, and thus, to anticipate reconfigurations (e.g., using TimeFlip~\cite{mizrahi2015timeflip}).

\subsection{CRR design}

In light of the above points, the problem we want to solve is to find the best assignment of $M$ TMs to $N$ robust RCs in order to find $N$ TM clusters having a minimum length of $L$ TMs and an overlap $O$. Moreover, since TMs' IDs are temporally ordered, the solution also provides the best expected cluster transition instants.

Note that the members of a cluster are required as input of the in-cluster robust routing, which, in turn is required to drive the TM clusters formation, through the estimated congestion. Therefore, this two aspects must be jointly addressed to obtain an optimal solution. Unfortunately, the problem is strongly combinatorial and a joint optimization model has revealed to be very hard to solve. State-of-the-art integer programming solvers, like Gurobi Solver\footnote{www.gurobi.com} or IBM CPLEX\footnote{www.ibm.com/software/commerce/optimization/cplex-optimizer/}, could not provide a solution in reasonable times: small instances of tens of TMs require several days to get the optimum.

\begin{figure}[t]
	\centering
\includegraphics[width=0.9\linewidth]{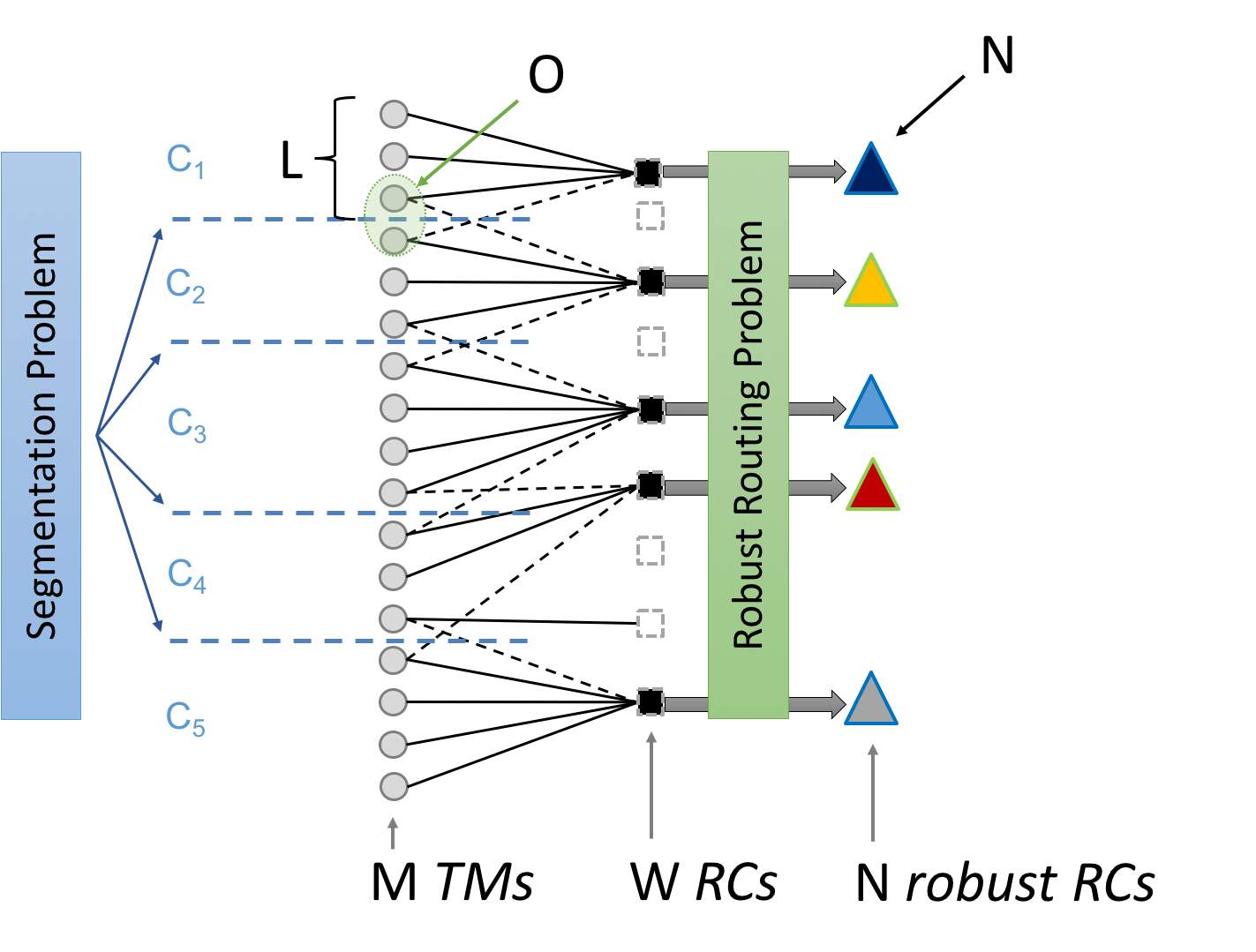}
	\caption{\small{High-level view of the proposed CRR algorithm.}}
    \vspace{-2em}
	\label{fig:crr_algo}
\end{figure}

The hardness of the joint problem calls for the development of a heuristic approach to split the overall complexity in more affordable subproblems. In this perspective, we propose the two-step Clustered Robust Routing (CRR) algorithm represented in Fig.~\ref{fig:crr_algo}. In the first step, a \textit{Segmentation Problem} is solved: the best assignment of $M$ TMs to $N$ out of $W$ given RCs is computed, considering the minimum holding time constraint and the overlap. In the second step, a \textit{Robust Routing Problem} is computed for each of the $N$ clusters, in order to create new RCs better customized for the selected TMs. The new RCs are introduced in the set of available RCs to the Segmentation Problem and the two steps are repeated for a given number of iterations.

%-------------------------------------------
% Segmentation Problem
%-------------------------------------------
\subsection{STEP 1 - Segmentation Problem}

The Segmentation Problem takes in input a set of TMs $\mathcal{T}=T(1),...,T(M)$ and a set of RCs $\mathcal{R}=R_1,...,R_W$. Its goal is to assign each TM $T(i)$ to a RC $R_j$ such that the overall association cost $\delta_{ij}$ is minimized and the number of used RCs is not larger than $N$. The cost $\delta_{ij}$ can be precomputed and corresponds to the network Maximum Link Utilization (MLU)\footnote{Note that we decided to used MLU as it is a commonly used metric that directly expresses the network congestion, however the model is general enough to consider other types of metric} when TM $T(i)$ is routed through RC $R_j$. This creates a set of $N$ TM clusters and RCs to manage the routing during the considered time period.

\begin{figure}
    \centering
    \begin{subfigure}[b]{0.3\columnwidth}
        \includegraphics[width=\textwidth]{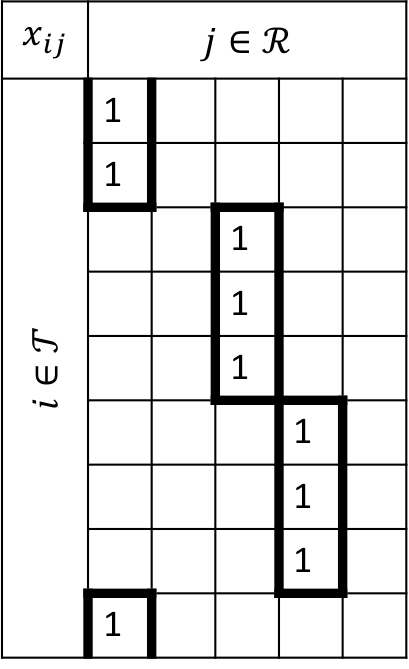}
        \caption{Segmentation with no overlap}
        \label{fig:segmentation_base}
    \end{subfigure}
    \hspace{0.5cm}
    \begin{subfigure}[b]{0.3\columnwidth}
        \includegraphics[width=\textwidth]{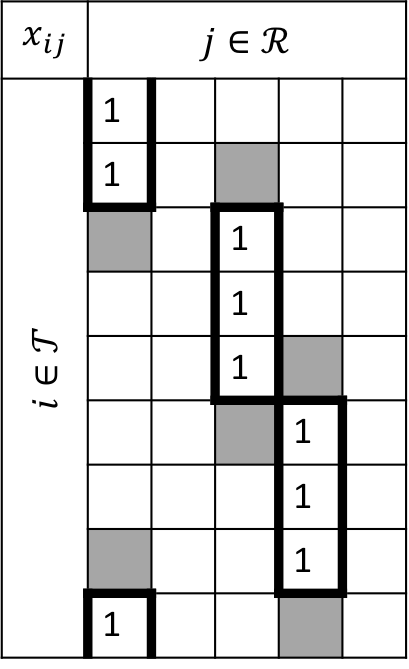}
        \caption{Segmentation with $O=1$}
        \label{fig:segmentation_overlap}
    \end{subfigure}
    \caption{Segmentation problem, $x_{ij}$ matrix}\label{fig:segmentation}
\end{figure}

We model the Segmentation Problem as an ILP model based on two sets on binary variables. Variables $x_{ij}, i \in \mathcal{T}, j \in \mathcal{R}$, setting the assignment of TM $T(i)$ to the cluster associated to RC $R_j$, and variables $z_{j}, j \in \mathcal{R}$, indicating with the value 1 that RC $j$ is used to form a cluster. If we visualize the matrix corresponding to variables $x_{ij}$ (see Fig.~\ref{fig:segmentation_base}), the solution of the problem is a set $N$ row-sequences of 1's with minimum length $L$. These sequences must be unique when appearing in a column, and correspond to a set of TM clusters associated to RCs. In order to identify the beginning of each cluster, we rely on variables $y_{ij}$, forward differences of variables $x_{ij}$. When $y_{ij}=1$, we can identify the initial TM of the cluster assigned to RC $R_j$: $T(i+1)$. The problem is fully described by the following ILP model:

%http://www.hostmath.com/
\begin{align}
    &\mathbf{[SP]:}\ \text{min.}\ \sum_{i\in \mathcal{T},j\in \mathcal{R}} x_{ij}\delta_{i,j} \qquad \text{s. t.:}\label{eq:base_of} \\
	&y_{ij} \geq x_{(i+1)_{|\mathcal{T}|}j} - x_{ij},\qquad \forall i \in \mathcal{T}, j \in \mathcal{R}\label{eq:diff}\\
	&\sum_{i\in \mathcal{T}} y_{ij} \leq z_{j},\qquad\forall j \in \mathcal{R}\label{eq:rc_activation}\\
	&\sum_{i\in \mathcal{T},j\in \mathcal{R}} y_{ij} \leq \sum_{j\in \mathcal{R}} z_{j} \label{eq:unicity}\\
	&\sum_{j\in \mathcal{R}} x_{ij} = 1,\qquad\forall i \in \mathcal{T}\label{eq:tm_assoc}\\
	&\sum_{i\in \mathcal{T}} x_{ij} \geq L \cdot z_{j},\qquad\forall j \in \mathcal{R}\label{eq:min_length}\\
	&\sum_{j\in \mathcal{R}} z_{j} \leq N\label{eq:rc_budget}\\
	&x_{ij},y_{ij},z_{j} \in \{0,1\},\qquad\forall i \in \mathcal{T},j \in \mathcal{R}
\end{align}
\noindent The objective function~(\ref{eq:base_of}) minimizes the sum of the association costs. The first set of constraints~(\ref{eq:diff})\footnote{The notation $(\cdot)_{m}$ indicates the modulo-$m$ operator} force variables $y_{ij}$ to be $1$ whenever the forward differences of variables $x_{ij}$ are $1$. Constraints~(\ref{eq:rc_activation}) force the activation of the variable $z_j$, associated to RC $R_j$, if a non-null forward difference is present in column $j$, which means cluster $C_j$ is considered in the solution. Constraint~(\ref{eq:unicity}) states that the number of non-null forward differences in the matrix must not exceed the number of selected RCs. Therefore, together with constraints~(\ref{eq:rc_activation}), this guarantees a unique cluster for each ``active'' column.
The set of constraints~(\ref{eq:tm_assoc}) impose each TM $T(i)$ to be assigned to a unique RC, while constraints~(\ref{eq:min_length}) force a minimum number of TMs associated to RC $R_j$, which, combined with the previous constraints imposing a unique and compact sequence of 1's within a column, correspond to constrain the minimum cluster length. Finally, constraint~(\ref{eq:rc_budget}) states that no more than $N$ clusters can be generated.

The set $\mathcal{R}$ is initialized by considering $W$ RCs obtained by the solution of the Robust Routing problem over $W$ sequential groups of TMs spanning the entire set $\mathcal{T}$. At the end of each algorithm iteration, the computed RCs are included in this initial set, this provides $\mathcal{R}$ with more refined RCs, which could be selected in the solution of the Segmentation Problem of the next iteration.

In order to consider an overlap between adjacent clusters, formulation $\mathbf{SP}$ must be amended to introduce the fact that up to $O$ TMs beyond the boundaries of the clusters could be routed with the RC associated to the cluster. We model this by stating that each of the $O$ TMs in overlap (grey cells in Fig.~\ref{fig:segmentation_overlap}) provides a congestion contribution $\delta_{ij}$ that is the average between the one of its associated cluster and the one of that in overlap\footnote{The model can capture other assumptions by simply changing some of the coefficients in the formulation.}. The following constraints:
\begin{align}
  	&w_{ij} \geq x_{(i-1)_{|\mathcal{T}|}j}-x_{ij},\qquad \forall i \in \mathcal{T}, j \in \mathcal{R}\label{eq:overlap_first}\\
	&\sum_{i\in \mathcal{T}} w_{ij} \leq z_{j},\qquad\forall j \in \mathcal{R}\label{eq:overlap_last}\\
	&\sum_{i\in \mathcal{T},j\in \mathcal{R}} w_{ij} \leq \sum_{j\in \mathcal{R}} z_{j}	
\end{align}
and a new objective function must be introduced in $\mathbf{SP}$:    
\begin{align}
    &\min ~~ \ \sum_{i\in \mathcal{T},j\in \mathcal{R}} x_{ij}\delta_{i,j}~+\nonumber\\
    &\frac{1}{2}\sum_{i\in \mathcal{T},j\in \mathcal{R}} y_{ij}\left(\sum_{(i-O < k \leq i)_{|\mathcal{T}|}}\delta_{k,j}-\sum_{(i+1\leq k\leq i+O)_{|\mathcal{T}|}}\delta_{k,j}\right) ~+\nonumber\\
    &\frac{1}{2} \sum_{i\in \mathcal{T},j\in \mathcal{R}} w_{ij} \left(\sum_{(i\leq k < i+O)_{|\mathcal{T}|}}\delta_{k,j}~-\sum_{(i-O\leq k < i)_{|\mathcal{T}|}}\delta_{k,j}\right)\label{eq:overlap_of}
\end{align}
Constraints~(\ref{eq:overlap_first})-(\ref{eq:overlap_last}) define variables $w_{ij}$ as backward differences of $x_{ij}$. Interpreted as the end of the compact row-sequences of 1's in Fig.~\ref{fig:segmentation}, $w_{ij}$ must satisfy the same uniqueness requirements as $y_{ij}$. The new objective function~(\ref{eq:overlap_of}) updates the association cost of TMs in overlap by removing half of the cost related to the associated cluster's RC and adding half of the cost towards the RC of the cluster in overlap.

%-------------------------------------------
% Robust Routing
%-------------------------------------------
\subsection{STEP 2 - Robust Routing Problem}

Once TMs have been clustered around an RC in STEP 1, STEP 2 computes a new robust RC $R$ considering the TMs in the cluster. This will likely provide a better customized routing. In addition, being a robust routing, it makes CRR intrinsically robust against noisy TM measurements.

We compute $R$ as robust RC that minimizes the MLU $\gamma_{max}$ measured over network links, $(i,j) \in \mathcal{L}$, when the set of TMs in cluster $C_c$, denoted as $\mathcal{T}_c$, is routed via $R$. The TMs in $\mathcal{T}_c$ are characterized by the same demand set $\mathcal{H}$, but different demand values, varying according to the traffic time evolution. We express the unique RC via flow variables $f_{ij}^{h}$, which indicate the amount of demand $h$ flow of every TM in $\mathcal{T}_c$ must be routed along the link $(i,j)$.

The ILP formulation of the Robust Routing Problem is:
\begin{align}
	&\mathbf{[RR]:}\ \text{min.} \ \gamma_{max} \qquad \text{s. t.:}\\
	&\sum_{(i,j)\in \mathcal{L}} f_{ij}^{h} - \sum_{(j,i)\in \mathcal{L}} f_{ji}^{h} =
	\begin{cases} 
		1 & \text{if } i = O_h\\
		-1 & \text{if } i = D_h\\
		0 & \text{otherwise}
	\end{cases}\nonumber\\
    &\hspace{1cm}\forall i \in \mathcal{N}, h \in \mathcal{H} \label{eq:routing}\\
    &\sum_{h\in \mathcal{H}} d_h^m \cdot f_{ij}^{h} \leq c_{ij}, \hfill \forall m \in \mathcal{T}_c, (i,j) \in \mathcal{L}\label{eq:capacity}\\
    &\gamma_{max}\geq \frac{\sum_{h \in \mathcal{H}} d_h^m f_{ij}^{h}}{c_{ij}} , \hfill\forall m \in \mathcal{T}_c, (i,j) \in \mathcal{L}\label{eq:mlu}\\
    &0\leq f_{ij}^{h} \leq 1, \qquad\forall h \in \mathcal{H}, (i,j) \in \mathcal{L}
\end{align}
Constraints~(\ref{eq:routing}) are standard flow conservation constraints for splittable routing\footnote{We consider here a more general splittable routing because it can be easily implemented on SDN switches, however the model can be easily modified to consider unsplittable routing solutions as well.}, with $O_h$ and $D_h$, respectively, origin and destination of the OD demand $h$. Constraints~(\ref{eq:capacity}) guarantee that the routing of each demand, with a request of $d_h^m$ units of flow in TM $T(m)$, does not exceed the link capacity $c_{ij}$ for any TM. Finally, constraints~(\ref{eq:mlu}), together with the objective function, implement a min-max of the standard link utilization formulation at RHS of (\ref{eq:mlu}) over the TMs in $\mathcal{T}_c$.
%-------------------------------------------
% Experimental/numerical results
%-------------------------------------------
\section{Numerical Results}
\label{results}

In order to assess the performance of the proposed algorithm, we consider a daily scenario in which we compare our CRR algorithm to different routing solutions within the Abilene Network \cite{abilene}, whose traffic requests are described by a set of TMs with granularity $5$ minutes ($288$ TMs for the entire day). We imagine a scenario in which clusters and RCs are optimized during the night for the day after, on the basis of daily TM predictions. Unless differently indicated, we average obtained results over a week and run the algorithm for $10$ iterations. The CRR algorithm has been implemented in Python, using Gurobi Solver language interface.

%-------------------------------------------
% Clustering approaches
%-------------------------------------------
\subsection{Clustering approaches comparison}
\begin{figure}[t]
	\centering
	\includegraphics[width=0.9\linewidth]{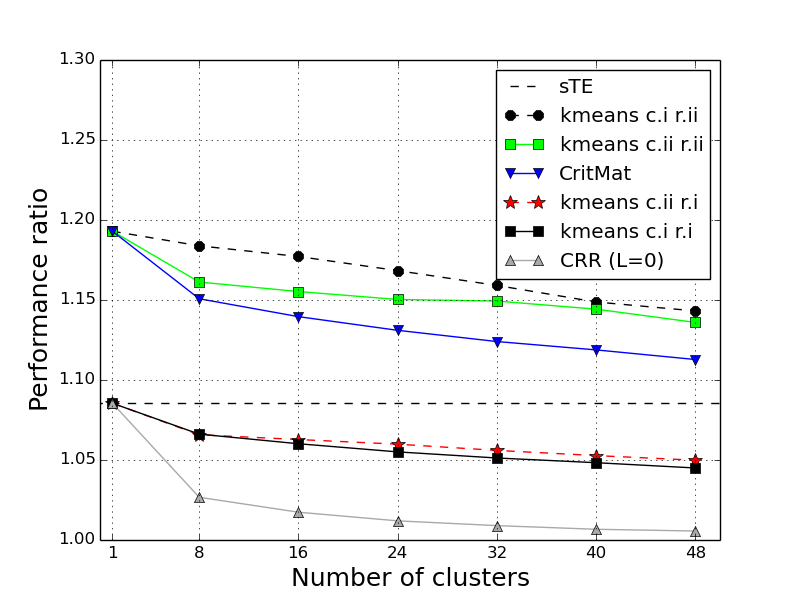}
	\caption{\small{Performance comparison of different TM clustering and robust routing approaches.}}
    \vspace{-1em}
	\label{fig:clustering_comparison}
\end{figure}

Fig.~\ref{fig:clustering_comparison} shows the comparison of different clustering techniques, for the moment we do not consider the minimum cluster length constraint $L$ and the overlap $O$. We measure the performance in terms of ratio of TM-averaged network MLU with respect to the value achievable by routing each TM through its MLU-optimum routing, that is, by applying dynamic TE. On the x-axis, the number of generated clusters is shown. We tested our CRR algorithm against different alternative approaches:
\begin{itemize}
\item \textit{sTE}: A static TE solution where a robust routing is computed over the entire TM set. The result is a single daily RC and no reconfiguration is required, like in the case of oblivious routing.
\item \textit{CritMat}: This approach, presented in \cite{zhang2005finding}, consists in clustering TMs according to dominating cluster heads, which are synthetic TMs including the maximum of each demand among the TMs grouped into the cluster. The RC associated to the cluster is the  MLU-optimum routing for the cluster head.
\item \textit{K-means clustering}: Most of the TM clustering approaches in literature are based on a variant of the well-known k-means technique. We have applied four k-means versions considering all combinations of the following clustering domains (\textit{c}) and in-cluster robust routing approaches (\textit{r}): \textit{c.i)} clustering in the TM domain (considering similarity among OD demand values) and \textit{c.ii)} clustering in the best-routing domain (considering similarity among MLU-optimum routing for each TM); \textit{r.i)} robust routing applied as in formulation $\mathbf{RR}$, \textit{r.ii)} MLU-optimum routing applied to the dominating TM of each cluster.
\end{itemize}
We can note how the proposed CRR algorithm outperforms all other alternatives. The curves' trend shows that CritMat dominating TM appears to be over-conservative, as the resulting congestion is even worse than that of static TE. Indeed, the outcome RCs can address such a large set of potential TMs that their working points are largely suboptimal when RCs are applied to specific TMs. The k-means approaches show very different congestion levels. The type of applied in-cluster robust routing is the main performance driver: $\mathbf{RR}$ formulation provides remarkably better results than relying on a dominating TM. The impact of the clustering approach, instead, is limited and its benefit depends on the type of robust routing strategy subsequently applied.

\begin{figure}[t]
	\centering
	\includegraphics[width=0.9\linewidth]{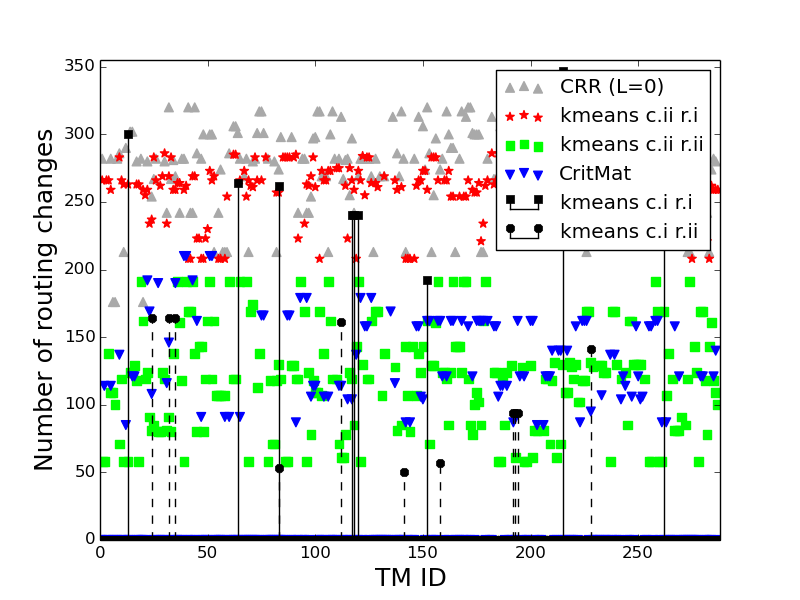}
	\caption{\small{Number of routing changes when different clustering approaches are applied. Example of solutions with $8$ clusters.}}
    \vspace{-1em}
	\label{fig:routing_changes}
\end{figure}

Fig.~\ref{fig:routing_changes} shows the reconfiguration intensity of the clustering approaches compared in Fig.~\ref{fig:clustering_comparison}. For the sake of clarity, we plot just the case in which $8$ clusters are generated. The figure considers a typical day, the time is expressed in terms of ordered TM IDs and each ID corresponds to a 5-minute interval. Each point in the plot indicates the number of links that change their routing coefficients ($f_{ij}^{h}$ variables in formulation $\mathbf{RR}$) with respect to the routing in the previous TM. If this value is $0$, it means that the previous RC is maintained. We can notice that CritMat and the k-means with routing-domain clustering produce many reconfigurations, frequently changing activated RCs. K-means with TM-domain clustering, instead, results more stable, however it still exhibits two main drawbacks. First, although being designed to use 8 clusters, it produces more reconfigurations (up to 10-11), as the 8 associated RCs are reused. Second, there are some reconfiguration bursts where RCs change after few minutes. The CRR algorithm with $L=0$, which provides the best performance in terms of congestion, is characterized by an unstable routing behavior as well. Therefore, we need to explicitly provide a minimum cluster length guarantee to avoid route flapping problems, which, as we will see, comes at the cost of a small congestion increase. This guarantee, however, results in a fixed number of points in Fig.~\ref{fig:routing_changes}, each separated the desired length $L$.

%-------------------------------------------
% L and O
%-------------------------------------------
\subsection{Impact of minimum cluster length and overlap}

\begin{figure}[t]
	\centering
	\includegraphics[width=0.9\linewidth]{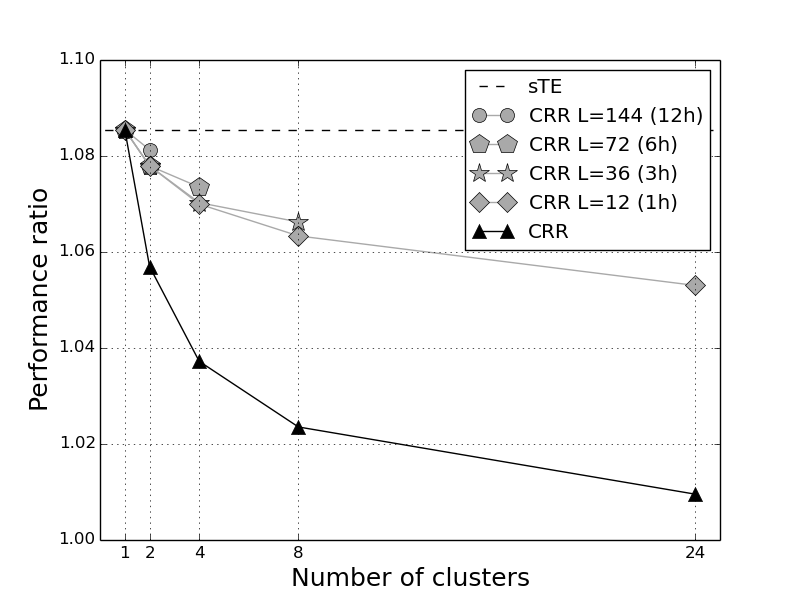}
	\caption{\small{Impact of different minimum cluster lengths on the CRR performance}}
    \vspace{-1.3em}
	\label{fig:clustering_mintime}
\end{figure}

In Fig.~\ref{fig:clustering_mintime}, we assess the performance of CRR algorithm when the minimum cluster length constraint is activated with different values of $L$. The x-axis shows the number $N$ of clusters in a day, while different curves represent different values of $L$. Note that the values of $L$ and $N$ are not independent, as $N$ clusters are generated in one day, $N$ cannot be larger than the ratio $24$ hours / $L$ (in hours). Therefore curves with larger $L$ stops at smaller $N$ values.  We can see that the minimum length constraint impacts on the performance of the clustering algorithm. With realistic $N$ values, the MLU performance ratio increases from values about $1.02$ (still referring to dynamic TE) to values about $1.06$. CRR with $N=8$ and $L=36$ results in keeping the same routing configuration for at least $3$ consecutive hours and changing only $8$ times the RC during the next day.

Focusing on $N=8$, the congestion of CRR with a minimum length $L=1h$ is $6.3$\% higher than that in dynamic TE, which is better than the performance of the closest alternatives with no time constraints, $6.5$\%, that of the TM-domain k-means clustering with robust routing. Therefore, our proposal, besides guaranteeing strong bounds on the level of reconfiguration, allows to even decrease the network congestion.

\begin{figure}[t]
	\centering
	\includegraphics[width=0.9\linewidth]{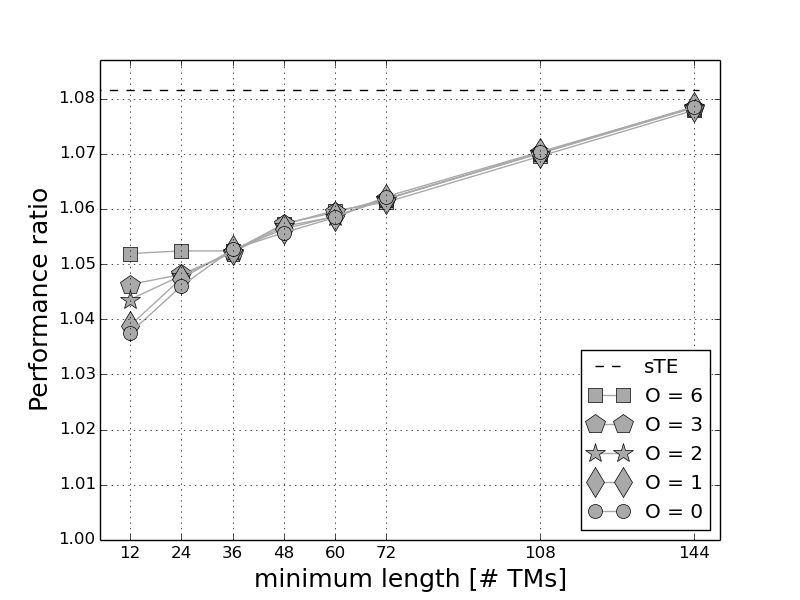}
	\caption{\small{Impact of different degrees of overlap on the CRR performance}}
    \vspace{-3.7em}
	\label{fig:clustering_overlap}
\end{figure}

In Fig.~\ref{fig:clustering_overlap}, we analyze the performance of the CRR algorithm varying the degree of overlap $O$. The figure shows on the x-axis the minimum length $L$ imposed to the cluster, while different curves are plotted for some values of $O$. We can see the impact of the overlap is significant only for short clusters, while it becomes quickly negligible when the minimum cluster length increases. Moreover, note that each TM included in the overlap provides an overlap extension of 5 minutes on each side. Therefore, considering $O=1,..,6$ means applying transition periods from $10$ minutes to $1$ hour, which we believe could reasonably include the complete set of meaningful values in practice, in terms of both uncertainness on the transition point and time required to anticipate reconfigurations.

%-------------------------------------------
% Prediction error
%-------------------------------------------
\subsection{Impact of prediction error}

%http://www.tablesgenerator.com/latex_tables#
\begin{table}[b]
\centering
\caption{\small{CRR performance when prediction error is considered. Results are expressed as percentage increase with respect to the ideal optimum routing.}}
\label{tab:prediction}
\begin{tabular}{|c|c||r|r|r|r|r|}
\hline
\multicolumn{2}{|r||}{\diagbox[width=2.5cm]{cluster.}{$\alpha$}}      & \multicolumn{1}{c|}{0} & \multicolumn{1}{c|}{15} & \multicolumn{1}{c|}{30} & \multicolumn{1}{c|}{45} & \multicolumn{1}{c|}{60} \\ \hline\hline
\multicolumn{2}{|c||}{sTE}   & 6.52                   & 7.02                   & 8.19                    & 9.25                   & \textbf{10.52}                   \\ \hline
\multirow{6}{*}{CRR} & L=72 & 4.07                   & 5.13                    & 6.93                   & \textbf{8.94}                    & 11.09                   \\ \cline{2-7} 
                     & L=60 & 4.04                   & 5.23                    & 7.24                    & 9.44                    & 11.19                   \\ \cline{2-7} 
                     & L=48 & 3.76                   & 5.06                   & 7.12                   & 9.61                    & 11.79                   \\ \cline{2-7} 
                     & L=36 & 3.17                   & 4.55                   & \textbf{6.74}                   & 9.04                   & 11.33                   \\ \cline{2-7} 
                     & L=24 & 2.84                   & 4.50                     & 6.85                    & 9.35                   & 11.93                   \\ \cline{2-7} 
                     & L=12 & \textbf{2.06}                   & \textbf{4.29}                   & 7.04                   & 10.08                  & 12.28                   \\ \hline
\end{tabular}
\end{table}

In the previous sections, we have analyzed the performance of the CRR algorithm when the clustering and the related RCs are computed over a set of  TMs and applied to the same set. This corresponds to assume perfect TM prediction and provide the potential performance achievable by the algorithm. In this section, we relax this assumption and analyze the impact of prediction errors. 

In order to reproduce the effect of unideal predictions, we run the CRR algorithm over a noisy version of the daily set of TMs to compute clusters and RCs, then apply the RCs to the original set of TMs, which represent the real traffic behavior. Each noisy TM version has been obtained from the original one by adding a uniform relative error $\left[-\alpha, \alpha \right] \%$ to every OD demand $d_h^m$. The results of these experiments are shown in Table \ref{tab:prediction}, where the performance achievable with different cluster lengths $L$ and prediction errors $\alpha$ is reported. Similarly to previous analyses, the performance is computed as the percentage increase of the average network MLU with respect to the ideal case of applying dynamic TE in perfect prediction conditions.

We can clearly note that the performance of CRR is negatively impacted by the presence of prediction error, however the intrinsic robustness of the clustered approach limits the performance decrease. Even with large errors, the gap with respect to the ideal Dynamic TE is within 10-12\%. The most interesting aspect to note is the parameters setting that provides the best performance, whose outcome is marked in bold in the table. The results show that the larger the error, the larger the clusters of the best solution. Indeed, when the quality of predicted TMs worsens, considering robust RCs computed over larger sets of TMs provides more robustness to any variation. A bigger variety of TMs included in the cluster used to generate a RC allows to better cope with traffic uncertainness. Taking this to extremes, when we have very low-quality predictions, no clustering can be helpful, because the representative set of TMs and the actual traffic will have little correlation. In these conditions, the Static TE approach, which builds a single RC considering all possible TMs in a day, is the best one can apply, as it generates the most robust RC. On the contrary, however, few and larger clusters lead to a bigger gap with respect to the dynamic TE, shown in Figg.~\ref{fig:clustering_comparison}~and~\ref{fig:clustering_mintime}, as the generated RCs are more conservative and far from being MLU-optimal for specific TMs. Therefore, a trade-off between cluster length and prediction accuracy exists. In case of good predictions, the size of the clusters drives the performance, vice versa, if predictions are affected by large errors, the impact of TM uncertainness completely overwhelms the effect of cluster sizes.

\subsection{Open issues}

This trade-off between cluster size and prediction accuracy opens a new technical challenge, which we cannot address in this paper, but appears to be a promising research direction. Thanks to the SDN paradigm, the controller can collect quasi-realtime measurements, thus can predict the traffic evolution and estimate a-posteriori the prediction error. This allows to anticipate the clusters that could be potentially visited in the near future and could provide the desired optimality gap with respect to an ideal congestion level. Moreover, the controller can prearrange a set of robust RCs derived from TM clusters, which, although referring to the same TM centroid, are characterized by different sizes, i.e., robustness levels, so that the system can easily shift through different RCs when the prediction accuracy suddenly changes. Finally, clusters can be synthetically generated as well, in order to include potentially severe anomalies RCs must be robust against.

The SDN controller must play the main role in managing the set of available RCs to orchestrate the routing of the entire SDN network over time by dispatching and activating the best RCs in each SDN switch. The design of the algorithm to select the type of generated RCs and the orchestration strategy is fundamental to provide performance optimality and full flexibility in front of traffic changes to advanced Software Defined Networks.

%-------------------------------------------
% Conclusion
%-------------------------------------------
\section{Conclusion}
\label{conclusion}

In this paper we investigated how robust routing approaches can be made adaptive in the SDN context. Assuming the availability of traffic predictions, we designed an off-line method to split the traffic space into smaller partitions and build routing configurations that are robust against any real-time traffic variation within the partition.

The results showed that routing configurations based on TM clustering can achieve a performance very close to the optimal routing only if a good clustering domain is chosen. Our proposal based on the estimation of the congestion caused by the activation of a given routing outperforms the other candidate solutions. This good performance is also confirmed when the clustering is further constrained by technical and practical issues on the obtained routing configurations, which we considered in our approach.

Finally, we investigated the behavior of our solution when the accuracy of traffic predictions varies. It showed an interesting trade-off between cluster sizes and prediction errors that opens a new research direction for the orchestration of robust routing configurations over time in SDN.

%-------------------------------------------
% References
%-------------------------------------------
\bibliographystyle{IEEEtran}

\end{document}